\documentclass[conference]{IEEEtran}
\usepackage{amsmath}
\usepackage{amssymb}
\usepackage{verbatim}
\usepackage{epsfig}
\usepackage{subfigure}
\usepackage{ifpdf}
\usepackage{cite}

\begin{document}
\title{Analysis of Fixed Outage Transmission Schemes:
A Finer Look at the Full Multiplexing Point}

\author{ \authorblockN{Peng Wu and Nihar Jindal}
\authorblockA{Department of Electrical and Computer Engineering \\
University of Minnesota \\
Email: pengwu, nihar@umn.edu}}
\maketitle

\begin{abstract}
This paper studies the performance of transmission schemes that
have rate that increases with average SNR while maintaining
a fixed outage probability.  This is in contrast to the classical
Zheng-Tse diversity-multiplexing tradeoff (DMT) that focuses on
increasing rate and decreasing outage probability.
Three different systems are explored: antenna diversity systems,
time/frequency diversity systems, and automatic repeat request
(ARQ) systems.  In order to accurately study performance
in the fixed outage setting, it is necesary to go beyond the
coarse, asymptotic multiplexing gain metric.  In the case of
antenna diversity and time/frequency diversity, an affine
approximation to high SNR outage capacity (i.e., multiplexing gain
plus a power/rate offset) accurately describes performance
and shows the very significant benefits of diversity.
ARQ is also seen to provide a significant performance advantage,
but even an affine approximation to outage capacity is unable to
capture this advantage and outage capacity must be directly
studied in the non-asymptotic regime.
\end{abstract}

\IEEEpeerreviewmaketitle
\section{Introduction}

In many time-varying communication systems, the receiver has
accurate \textit{instantaneous} channel state information (CSI),
generally acquired from received pilot symbols, while the
transmitter only knows the channel statistics (e.g., average
received SNR and the fading distribution) but has no instantaneous
CSI. This could be the case if, for example, the channel coherence
time is long enough to allow for receiver training (over a
reasonably small fraction of the coherence time) but this
information cannot be fed back to the transmitter because the
feedback delay is too large relative to the coherence time.

Performance in such a setting is generally dictated by fading
and the relevant performance metric is known to be the \textit{outage
probability}, which is the probability that the instantaneous mutual
information is smaller than the transmission rate, because this
quantity reasonably approximates the probability of decoder (frame)
error if a strong channel code is used \cite{pc}.  Such systems have traditionally been studied by considering
outage probability versus average SNR for a \textit{fixed}
transmission rate, leading to measures such as diversity order
(generally defined as the slope of the outage vs. SNR curve in
log-scale).

In modern communication systems, however, transmission rate is
generally adjusted according to the average SNR (via adaptive modulation and coding)
and thus systems
need to be studied at various rates and SNR levels. The seminal work
of Zheng and Tse \cite{pa} took precisely this viewpoint in
introducing the diversity-multiplexing tradeoff (DMT). Loosely
speaking, the DMT framework considers the performance of a
\textit{family} of codes indexed by average SNR such that the coding
rate increases as $r \log_2 SNR$, and the outage/error probability of
the code decreases approximately as $SNR^{-d}$. The quantity $r$ is
the \textit{multiplexing gain} while $d$ is the \textit{diversity
order} (of the family of codes, not of a particular code).  The DMT
region is the set of $(r,d)$ pairs achievable by \textit{any} family
of codes, and can be simply quantified in terms of the number of
transmit and receive antennas, $N_t$ and $N_r$ respectively, and the
receiver strategy for MIMO systems.

Over the past few years the DMT framework has become the benchmark for
comparing different transmission strategies for different systems
(MIMO, cooperative transmission, multiple access channel).  Although
the DMT framework has been incredibly useful in this role by
providing a meaningful and tractable metric to compare different
schemes that simultaneously achieve \textit{increasing rate}
and \textit{decreasing outage probability},
one very important paradigm not sufficiently captured by the DMT are
codes that achieve \textit{increasing rate} and \textit{fixed outage probability}.

Families of codes that achieve a fixed rather than decreasing outage are
important because they are used in a number of
important wireless systems, most prominently in the cellular domain.
In this setting, as the average SNR of a user increases (i.e., as a
user moves closer to the base station), it is desirable to use the
additional SNR to increase rate but not to decrease
outage probability (i.e., packet error rate); indeed, many systems
continually adjust rate precisely to maintain a target error probability
(e.g., $10^{-2}$).
In a voice system, for example, the voice decoder
may be able to provide sufficient quality if no more than $1\%$ of packets are
received incorrectly and thus there is no benefit
to decrease outage below $1\%$.  Therefore, serving each user
at the largest rate that maintains $1\%$ outage minimizes
per-user resource consumption (i.e., time/frequency slots)
and thereby maximizes system capacity.

In order to accurately study fixed-outage schemes, it is necessary
to directly study the manner in which outage capacity scales with
SNR. We denote outage capacity as $R(P, \epsilon)$, where this
quantity is the rate that achieves an outage probability of
$\epsilon$ at an average SNR of $P$. In the context of the DMT,
fixed outage systems (for any $\epsilon > 0$) achieve zero diversity
($d=0$) and thus can achieve the maximum multiplexing gain.  In
other words, the DMT tells us that $R(P, \epsilon) \approx
r_{\textrm{max}} \log_2 P$ for any $\epsilon > 0$, where
$r_{\textrm{max}}$ is the maximum multiplexing gain, but cannot
provide a more precise characterization than multiplexing gain (or
pre-log factor). In many scenarios of interest, such a
characterization is not sufficient to meaningfully characterize
performance.

\begin{figure}
\label{fig-dmt_outage}
\begin{center}
\subfigure[DMT Regions]{\label{fig:divcomp}
\includegraphics[width = 1.6in]{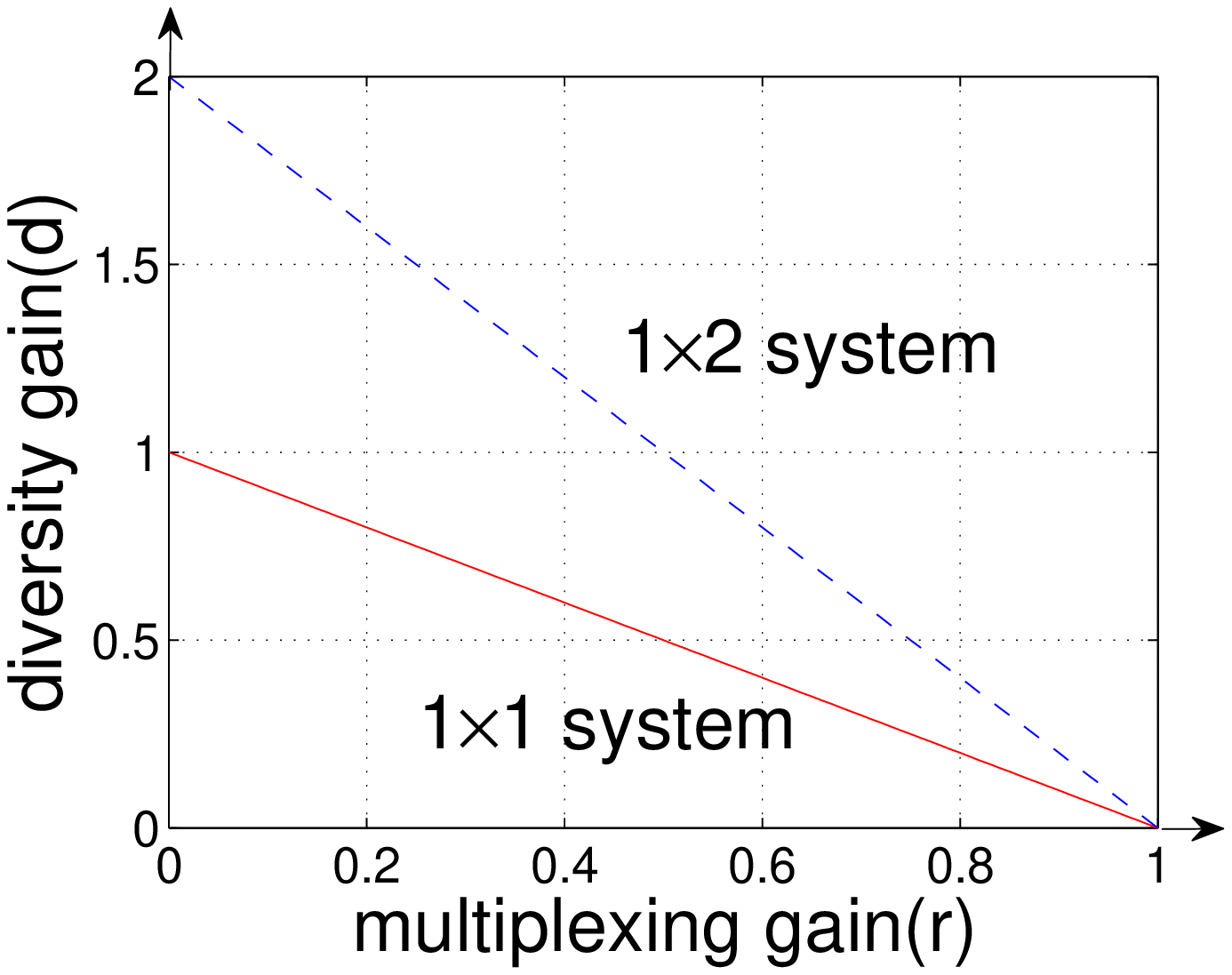}}
\subfigure[$R(P, \epsilon)$ for $\epsilon = .01$]{\label{fig:fullmuxcomp}
\includegraphics[width = 1.6in]{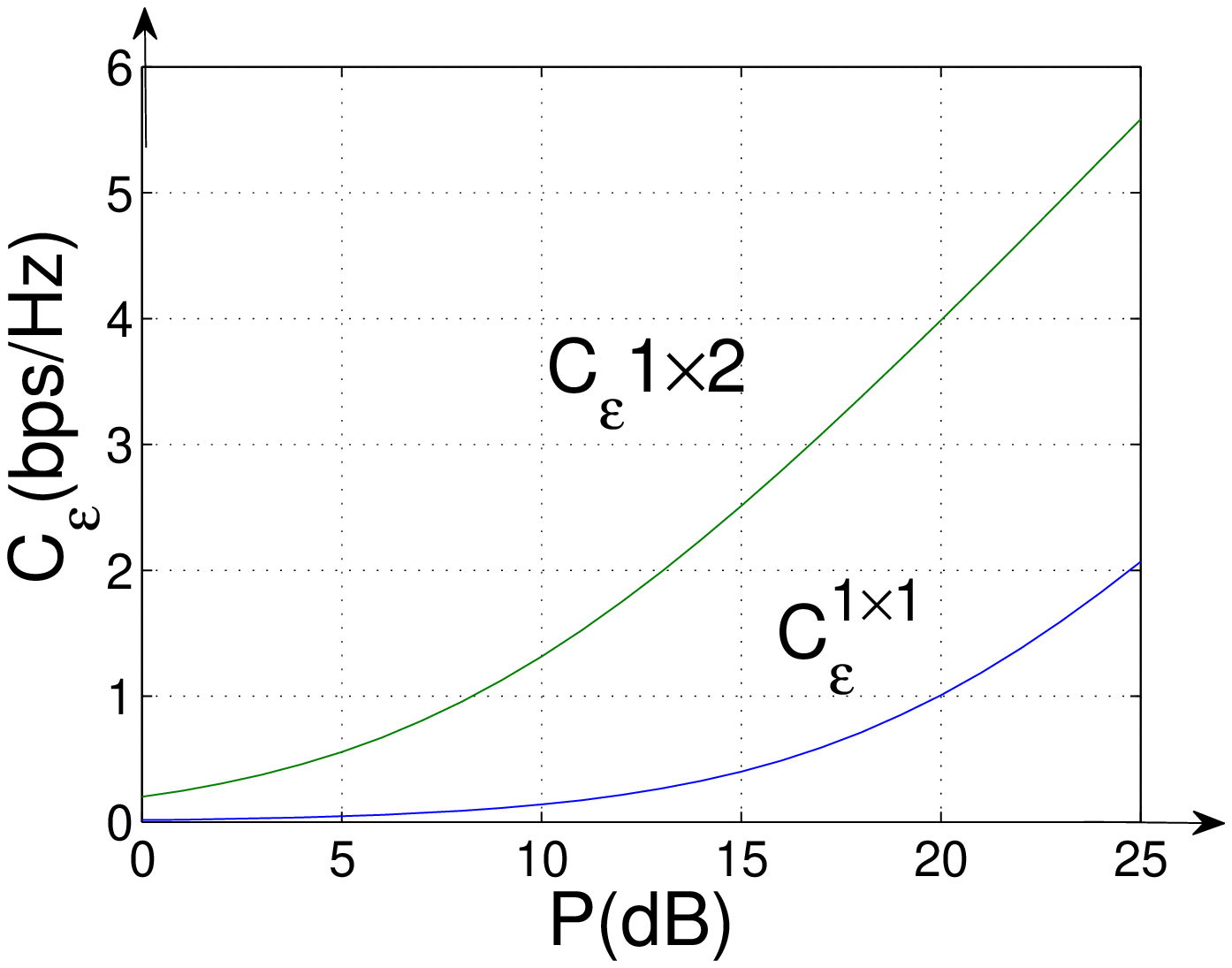}}
\caption{Full multiplexing performance in the the diversity-multiplexing tradeoff}
\end{center}
\end{figure}

To further illustrate the need to directly study outage capacity, let us consider a
simple example. For a $1 \times 1$ system the maximum diversity order
is $d^*(r)=1-r$, while for a $1 \times 2$ ($N_t=1$, $N_r=2$) system
$d^*(r)= 2(1-r)$ \cite{pa}.  The DMT regions for both systems are shown in Fig.
1 (a).  Because $\min(N_t,N_r)=1$ both regions share the
same full multiplexing point ($r=1, d=0$), a DMT-based comparison would indicate
that the systems are equivalent in a fixed outage setting.
However, the plot of $R(P, \epsilon)$ for $\epsilon = .01$ in
Fig. 1 (b) shows that there is a huge
power gap ($11.68$ dB) between the two systems; it clearly is not
sufficient to consider only the multiplexing gain, which is
the slope of the $R(P, \epsilon)$ curve.

Motivated by this example, one step in the right
direction is to consider \textit{affine} rather than \textit{linear}
approximations to outage capacity (at high SNR):
\begin{eqnarray} \label{eq-affine_outage}
R(P, \epsilon) = r_{\textrm{max}} \log_2 P + O(1),
\end{eqnarray}
where the non-vanishing $O(1)$ term, which depends on $\epsilon$ and the system configuration
(i.e., $N_t$ and $N_r$), is capable of capturing power/rate offsets
such as that seen in Fig. 1 (b).  This affine
approximation, first proposed in \cite{ph}, has been
extremely useful in analysis of the ergodic capacity of
CDMA systems  \cite{ph} and MIMO systems \cite{pi} \cite{pj}.
More recently, the affine approximation has also been employed to study the
outage capacity of MIMO systems at asymptotically high SNR \cite{pf}.
In \cite{pf}, an expression for the constant term in (\ref{eq-affine_outage})
is given in terms of the statistics of the channel matrix (more precisely,
in terms of the distribution of the determinant of ${\bf H}{\bf H}^H$ where
${\bf H}$ is the channel matrix).

\subsection{Contribution of Work}
In this paper, we first consider the case of antenna diversity (SIMO
or MISO; Section \ref{sec-antenna}) and show that fixed outage
capacity can be exactly specified in terms of the inverse of the
fading CDF.  Although this result can be viewed as a special case of
\cite{pf} (and also appears in \cite[Section 5.4]{TseVis}), it is
useful to consider this base case to more precisely illustrate the
importance of fixed-outage analysis. Next we consider systems with
time and/or frequency diversity (Section \ref{sec-timefre}), which
are modeled as block-fading channels, again in the context of the
affine approximation to outage capacity. Finally, we consider the
performance of systems using hybrid ARQ (automatic repeat request)
for incremental redundancy as well as chase combining. Unlike
antenna or time/frequency diversity, the benefit of H-ARQ vanishes
at asymptotically high SNR; in other words, the high-SNR affine
approximation to outage capacity is unaffected by H-ARQ.  However,
H-ARQ does provide a very significant advantage of low and moderate
SNR's.  In order to understand these gains, we directly study outage
capacity at finite SNR's.

\section{System Model}\label{sec-model}
We consider a block-fading channel, denoted by ${\bf H}$, which is randomly
drawn according to a known probability distribution
(e.g., spatially white Rayleigh fading) and then fixed for the duration of
a codeword.  Furthermore, the receiver is assumed to have perfect
channel state information (CSI) but the transmitter has no instantaneous
knowledge of the channel realization and is only aware of the probability
distribution.  The received signal ${\bf y}$ is given by:
\begin{eqnarray}
{\bf y} = \sqrt{P} {\bf H} {\bf x} + {\bf z},
\end{eqnarray}
where the input ${\bf x}$ is constrained to have unit norm, ${\bf z}$ is the complex Gaussian noise and $P$ represents the (average) received SNR.  We consider cases where the
input is Gaussian and further specify its structure where needed.

The outage probability is the probability that the
\textit{instantaneous mutual information} is smaller than the transmission rate $R$:
\begin{eqnarray}
P_{out}(R, P) = {\mathcal P} [ I(X;Y | {\bf H}, P) < R],
\end{eqnarray}
and the outage capacity $R(P,\epsilon)$ is defined as the maximum rate
that achieves the desired outage probability:
\begin{eqnarray}
R(P,\epsilon) \triangleq \sup_{ P_{out}(R, P) \leq \epsilon} R.
\end{eqnarray}
Note that this quantity is essentially the same as the $\epsilon$-capacity defined
by Verdu and Han \cite{pb}\footnote{In some cases we compute outage probability assuming
the input ${\bf x}$ is Gaussian and spatially white, while the precise
definition of $\epsilon$-capacity requires an explicit optimization over the
input distribution. This choice of input is easily shown to be optimal when
$N_t = 1$, but is not necessarily optimal for $N_t > 1$.}.

\section{Antenna Diversity} \label{sec-antenna}

We begin by examining antenna diversity, which is one of the commonly employed
diversity techniques.  The results of this section are a special
case of \cite{pf}, and precisely the same analysis appears
in \cite[Section 5.4]{TseVis}; thus this section should be treated as background
material.

If the transmitter has $N_t>1$ antennas while $N_r =1$, the mutual information
for a spatially white Gaussian input (components of ${\bf x}$ are iid Gaussian
with variance $\frac{1}{N_t}$) is $\log_2\left(1 + ||{\bf H}||^2 \frac{P}{N_t}\right)$
and therefore the outage probability is given by:
\begin{eqnarray}
P_{out}(R, P) = {\mathcal P} \left[
\log_2\left(1 + ||{\bf H}||^2 \frac{P}{N_t}\right) < R \right].
\end{eqnarray}
By setting this quantity to $\epsilon$ and solving for $R$, we get:
\begin{eqnarray}
C_{\epsilon}^{N_t \times 1}(P) & = & \log_2\left(1 + F_{||{\bf H}||^2}^{-1}(\epsilon) \frac{P}{N_t}\right),
\end{eqnarray}
where $F_{||{\bf H}||^2}^{-1}(\cdot)$ is the inverse of the CDF of random variable
$||{\bf H}||^2$.
In iid Rayleigh fading the components of ${\bf H}$ are iid $\mathcal{CN}(0,1)$ and
thus $||{\bf H}||^2$ is chi-square with $2N_t$ degrees of freedom and has
the following CDF:
\begin{eqnarray}
F_{||{\bf H}||^2}(x)
& = & 1 - e^{-x}\sum_{k=1}^{N_t} \frac{x^{k-1}}{(k-1)!}.
\end{eqnarray}
If $N_t =1$ the channel gain is exponential and the inverse CDF can be written in
closed form to yield:
\begin{eqnarray} \label{eq-1by1}
C_{\epsilon}^{1 \times 1}(P) & = & \log_2\left(1 +
\ln\left( \frac{1}{1-\epsilon} \right) P\right),
\end{eqnarray}
while for $N_t>1$ the inverse CDF needs to be computed numerically.

It can be convenient to relate the outage capacity to the AWGN capacity at SNR $P$:
$C_{AWGN}(P) = \log_2(1+P)$:
\begin{eqnarray}
C_{\epsilon}(P) =  C_{AWGN}(\Gamma_{\epsilon} P) =
\log_2(1+\Gamma_{\epsilon} P)
\end{eqnarray}
where the gap to capacity is:
\begin{eqnarray}
\Gamma_{\epsilon}^{N_t \times 1} =  \frac{F_{||{\bf H}||^2}^{-1}(\epsilon)}{N_t}.
\end{eqnarray}
For small $\epsilon$ we can approximate the CDF of $||{\bf H}||^2$
as $F_{||{\bf H}||^2}(x) \approx \frac{x^{N_t}}{N_t!}$
and therefore the gap can be approximated as:
\begin{equation}
\Gamma_{\epsilon}^{N_t \times 1}  \approx  \epsilon^{\frac{1}{N_t}}
\frac{ (N_t!)^{\frac{1}{N_t}}}{N_t}.
\end{equation}

In the case of receive diversity ($N_t=1, N_r > 1$) the achieved mutual information
is $\log_2\left(1 + ||{\bf H}||^2 P \right)$, because using optimal maximum-ratio
transmission prevents the power loss experienced with transmit diversity, and
therefore:
\begin{eqnarray}
C_{\epsilon}^{1 \times N_r}(P) & = & \log_2\left(1 + F_{||{\bf H}||^2}^{-1}(\epsilon) P \right),
\end{eqnarray}
where  $||{\bf H}||^2$ is chi-square with $2N_r$ degrees of freedom.
As expected, there is a $\log_{10}(N_t)$ dB power gap between $C_{\epsilon}^{1 \times N_r}(P)$
and $C_{\epsilon}^{N_t\times 1}(P)$.

In Fig. 2 the outage capacity of
$2 \times 1$, $1 \times 2$, and $1 \times 1$ systems are plotted for
$\epsilon = 0.01$.  The capacity gap for the $1 \times 1$ system
is approximately 20 dB ($\Gamma_{\epsilon} = - \ln(1-\epsilon) \approx \epsilon$), while it is
about 11.5 dB for the $2 \times 1$ system ($\Gamma_{\epsilon} \approx \sqrt{\frac{\epsilon}{2}}$).
Fixed outage analysis very clearly illustrates the advantage of antenna diversity.
\label{antenna_capacity}
\begin{figure}[ht]
\begin{center}
\includegraphics[width = 2.5in]{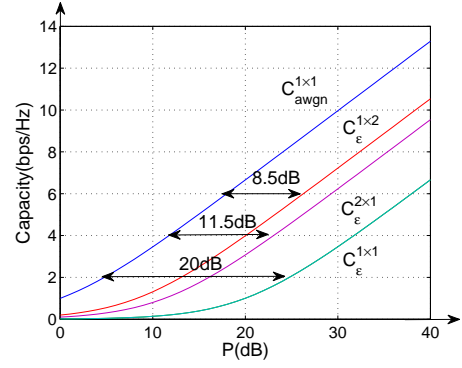}
\caption{$C_{awgn}$ and $C_{\epsilon}$ (bps/Hz) vs
$P$ (dB),$\epsilon=0.01$}
\label{fig:antenna_capacity}
\end{center}
\end{figure}

\section{Time / Frequency Diversity}\label{sec-timefre}

Another very common method of realizing diversity is through time or frequency, i.e.,
by coding across multiple coherence times/bands.  If a codeword spans $L$
coherence bands (in time and/or frequency), the outage probability is given by \cite[Equation 5.83]{TseVis}:
\begin{eqnarray} \label{eq-outage_time}
P_{out}(R, P) = {\mathcal P} \left[\frac{1}{L}\sum_{i=1}^{L}\log_2(1+P|h_i|^2)<R \right]
\end{eqnarray}
where $h_i$ is the channel gain over the $i$-th band, each $h_i$ is unit variance
complex Gaussian (Rayleigh fading), and $h_1, \ldots, h_L$ are assumed to be iid.
It is important to note that this outage probability expression approximates
the performance of a strong channel code that is interleaved across the
$L$ bands, and not that of a sub-optimal repetition code.

For notational convenience we define the function $G_L(R)$ to be equal to the
outage expression in (\ref{eq-outage_time}).  In terms of this function
\begin{eqnarray}
C_{\epsilon}(P) = G_L^{-1}(\epsilon).
\end{eqnarray}
Although we cannot reach a closed form for $G_L^{-1}(\epsilon)$, this quantity
can be computed numerically by noting that $C_{\epsilon}(P)$ is equal to the
$R$ that satisfies:
\begin{eqnarray}
\epsilon = \int\int\cdots\int_{\frac{1}{L}\sum_{i=1}^L \log_2(1 + x_i) < R} \frac{1}{P^L}e^{-\frac{x_1+x_2+\cdots\ x_L}{P}} \nonumber\\
dx_1dx_2\cdots dx_L.
\end{eqnarray}

A simple application of Jensen's inequality shows that the mutual information
achieved with $L$-order time/frequency diversity is smaller than that
achieved in a $L \times 1$ system:
\begin{eqnarray} \label{eq-outage_time}
\frac{1}{L}\sum_{i=1}^{L}\log_2(1+P|h_i|^2) \leq \log_2 \left (1+ \frac{P}{L}
\sum_{i=1}^L  |h_i|^2 \right).
\end{eqnarray}
As a result, the outage probability is larger for time/frequency diversity
and therefore the outage capacity of a $L \times 1$ system is no smaller than
the outage capacity of an $L$-order time/frequency diversity system.
\begin{figure}[ht]
\begin{center}
\includegraphics[width = 2.5in]{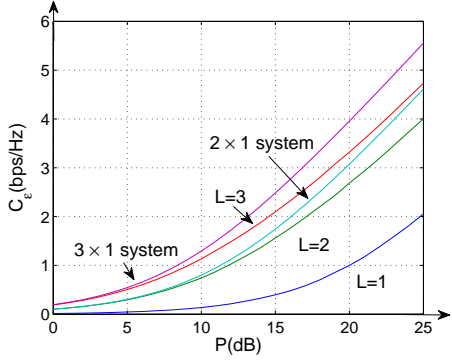}
\caption{Outage capacity of the time/frequency diversity system and the transmit diversity system,$\epsilon = 0.01$}
\label{fig:FreDivTransDiv}
\end{center}
\end{figure}
In Fig. 3, outage capacity is shown for $L=1,~2$ and $3$ along with
the outage capacity of $2 \times 1$ and $3 \times 1$ systems for $\epsilon = .01$.
The time/frequency diversity curve is smaller than the corresponding antenna
diversity system, but the difference is relatively small for low and moderate SNR. However, there is a nontrivial gap at high SNR that can be explained by the concavity of the $\log$ function. Note that there is
again a very significant gap between diversity order one ($L=1$) and two ($L=2$).

\section{ARQ} \label{sec-arq}

ARQ protocols can significantly improve performance by
allowing for retransmission of packets or transmission of additional parity
symbols when an initial transmission is unsuccessful.  We are particularly
interested in the performance of \textit{hybrid ARQ} (H-ARQ) protocols that allow
for decoding on the basis of multiple received packets.
Upon reception of a packet, the receiver attempts to decode and feeds back
a one-bit ACK/NACK message (often based on the success or failure of a CRC
check).  If an ACK is received the transmitter moves on to the next message, while
a NACK results in retransmission of the same packet (chase combining)
or transmission of additional parity symbols (incremental redundancy).
There generally is a limit to the number of ARQ rounds for the same message, denoted by $L$,
and an outage occurs whenever the message cannot be successfully decoded after
$L$ ARQ rounds.

If each message contains $b$ bits and each transmitted packet is $T$ symbols
long, then the initial rate of transmission is $R \triangleq \frac{b}{T}$.
If random variable $X$ is used to denote the number of ARQ rounds used for a
particular message (clearly $X \leq L$), then the long-term \textit{transmitted
rate} is \cite{CaireTuninetti}:
\begin{equation} \label{eq-defn_arq_rate}
\eta = \frac{R}{E[X]}.
\end{equation}
To see why this is the case, note that the number of packet transmissions used
to attempt to transmit $N$ messages is $\sum_{i=1}^N X_i$, where $X_i$ is
the number of ARQ rounds used for the $i$-th message.  Thus, the average
transmission rate (in bits/channel symbol) is:
\begin{equation}
\frac{N b} {T \sum_{i=1}^N X_i} = \frac{R} {\frac{1}{N} \sum_{i=1}^N X_i},
\end{equation}
and this quantity converges (by the law of large numbers) to
$\frac{R}{E[X]}$ as $N \rightarrow \infty$.

The choice of H-ARQ strategy and the outage constraint $\epsilon$
determine the initial rate $R$ and the distribution of $X$, and in
the following sections we analyze the performance of both incremental
redundancy (IR) and chase combining (CC).  We consider the case
where the channel in each ARQ round is independently drawn.

\subsection{Incremental Redundancy}

We first investigate incremental redundancy techniques, in which the the transmitter
sends additional parity bits (rather than retransmitting the same packet) whenever
a NACK is received.  In this setting it has been shown that the total
mutual information is the \textit{sum} of the mutual information received in each
ARQ round, and that decoding is possible once the \textit{accumulated} mutual
information is larger than the number of information bits \cite{CaireTuninetti}.
In other words, the number of ARQ rounds $X$ is the smallest number $l$ such that:
\begin{eqnarray}
\sum_{i=1}^l \log_2(1+P|h_i|^2) > R.
\end{eqnarray}
The constraint caps this quantity by $L$, and an outage occurs
whenever the mutual information after $L$ rounds is smaller than $R$:
\begin{eqnarray} \label{eq-ir_outgae}
P_{out}(R) = { \mathcal P} \left[\sum_{i=1}^L \log_2(1+P|h_i|^2) <R \right].
\end{eqnarray}
For simplicity we consider single antenna systems ($N_t=N_r=1$), and
use $h_i$ to denote the channel during the $i$-th ARQ round. In the
following sections we consider the case where the
channel is iid across ARQ rounds.
Similar to the definition in Section \ref{sec-timefre}, here we use $G_{IR,i}(R)$ to denote the probability that the sum of mutual information is less than the first round rate $R$ after $i$ rounds. Then,
$R = G_{IR,L}^{-1}(\epsilon)$.
Go back to the definition of $\eta$, we have
\begin{eqnarray}
\eta_{IR} & = & C_{\epsilon}^{IR,L} = \frac{R}{E[X]}\nonumber\\
     & = & \frac{G_{IR,L}^{-1}(\epsilon)}{1+\sum_{i=1}^{L-1}G_{IR,i}(G_{IR,L}^{-1}(\epsilon))}
\end{eqnarray}

It is useful to compare performance to a system without ARQ that
always codes over the $L$ available slots (whereas ARQ allows for
early completion), which precisely corresponds to $L$-order
time/freq diversity (Section \ref{sec-timefre}).  After properly
normalizing rates, we get:
\begin{equation}
\frac{C_{\epsilon}^{IR,L}}{C_{\epsilon}^{nARQ}} = \frac{L}{E[X]}
\end{equation}
where $C_{\epsilon}^{nARQ}$ is the outage capacity of a corresponding no ARQ protocol. Since $L\geq E[X]$, then
\begin{equation}
C_{\epsilon}^{IR,L} \geq C_{\epsilon}^{nARQ}
\end{equation}
Actually, the quantity $\frac{L}{E[X]}$ determines the advantage of ARQ, and it is not difficult to show the following limit:
\begin{equation}
\lim_{P\rightarrow \infty} E[X] =  L
\end{equation}
This shows that the effect of ARQ vanishes at asymptotically high
SNR because the expected number of ARQ rounds converges to the
maximum L. Indeed, it can further be shown that the rate advantage
of ARQ also vanishes at asymptotically high SNR:
\begin{equation}
\lim_{P\rightarrow \infty} [C_{\epsilon}^{IR,L}(P) - C_{\epsilon}^{nARQ}(P)] =  0
\end{equation}
In other words, the high SNR affine approximation is the same regardless of whether ARQ is used.
\label{ARQnARQ}
\begin{figure}[t]
\begin{center}
\includegraphics[width = 2.5in]{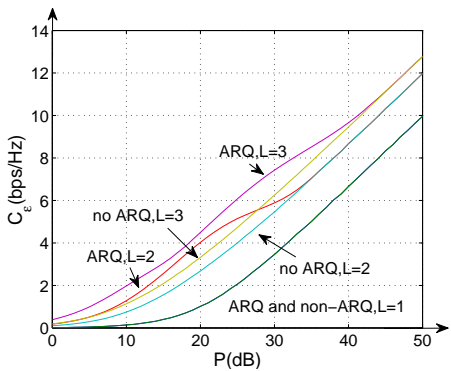}
\caption{Outage capacity for the IR strategy and the no ARQ strategy in the iid Rayleigh fading channel,$\epsilon = 0.01$}
\label{fig:IRARQnARQ}
\end{center}
\end{figure}
On the other hand, the number of expected ARQ rounds is less than
$L$ at asymptotically low SNR.

Based on these asymptotic results one might conclude that ARQ
provides a benefit only at relatively low SNR's. However, numerical
results indicate that the high SNR asymptotics kick in only for very
large SNR's. Indeed, ARQ does achieve a significant advantage for a
relatively large range of SNR's. In Fig. 4 the outage capacity is
shown for $\epsilon = 0.01$ and $L=1,2$ and $3$. Note that 2 rounds
of ARQ provide a significant power advantage relative to no ARQ up
to approximately 30 dB, while the advantage lasts past 40 dB for
$L=3$. Asymptotic measures such as multiplexing gain and rate/power
offset are clearly misleading in this context.

\subsection{Chase Combining}
If chase combining is used, the transmitter retransmits the same
packet whenever a NACK is received and the receiver performs
maximal-ratio-combining (MRC) on all received packets.. As a result,
SNR rather the mutual information is accumulated over ARQ rounds and
the outage probability is given by:
\begin{eqnarray}
P_{out}(R) = {\mathcal P}\left[\log_2(1+P\sum_{i=1}^L|h_i|^2)<R \right]
\end{eqnarray}
Note that this strategy essentially allows a repetition code to be used up to $L$ times. A straightforward derivation shows the outage capacity of CC in the iid Rayleigh fading channel is:
\begin{equation}
C_{\epsilon}^{CC,L}(P) = \frac{R}{1+(1-e^{-\frac{2^R-1}{P}})\sum_{k=1}^{L-1}(L-k)\frac{(2^R-1)^{k-1}}{P(k-1)!}}
\end{equation}
where $R$ has to be obtained from $G_{CC,L}^{-1}(\epsilon)$ numerically. Chase combining indeed provides some advantage at low and moderate SNR, but performs poorly at high SNR because of the sub-optimality of the repetition codes. In Fig. 5 we compare the performance of IR, CC and no ARQ strategy for $L=4$. We see that CC performance reasonably at low SNR but trails off at high SNR.  
\label{CCARQnARQ}
\begin{figure}[t]
\begin{center}
\includegraphics[width = 2.5in]{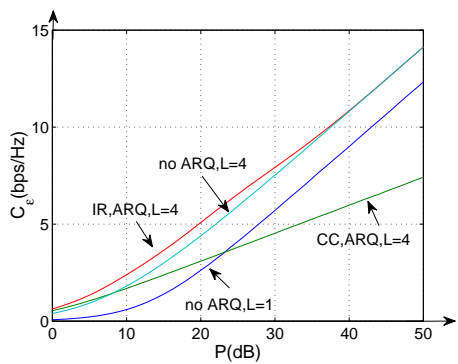}
\caption{Comparison of the outage capacity among IR strategy, CC strategy and no ARQ strategy,$\epsilon = 0.05$}
\label{fig:CCARQnARQ}
\end{center}
\end{figure}

\section{Conclusion}

In this paper we have studied open-loop communication systems under
the assumption that rate is adjusted such that a fixed outage
probability is maintained. The over-arching takeaways of this work
are two-fold. First, we have argued that schemes that increase rate
but have a \textit{fixed} rather than decreasing outage probability
may be more practically relevant than the increasing rate/decreasing
outage schemes addressed by the diversity-multiplexing tradeoff.
Second, we have shown that asymptotic measures should be used very
carefully in analysis of fixed-outage systems.  Multiplexing gain is
certainly too coarse in this context, while high SNR rate/power
offsets are sometimes meaningful (antenna diversity, time/frequency
diversity) but can also be misleading in other settings (e.g., ARQ
systems) due to their asymptotic nature.

We hope this paper establishes a more meaningful and practically
relevant framework by which different communication techniques, such
as partial channel feedback and relaying, can be studied by the
research community.

\end{document}